\documentclass[%
 reprint,
superscriptaddress,
showpacs,preprintnumbers,
nofootinbib,
nobibnotes,
 amsmath,amssymb,
aps,
pra,
floatfix,
]{revtex4-1}

\usepackage{graphicx}% Include figure files
\usepackage{dcolumn}% Align table columns on decimal point
\usepackage{bm}% bold math
\usepackage{hyperref}% add hypertext capabilities
\usepackage{tabularx}
\usepackage{amsmath}

\usepackage{mwe}    % loads »blindtext« and »graphicx«
\usepackage[caption=false]{subfig}
\usepackage{subfloat}

\usepackage[utf8]{inputenc}

\usepackage{natbib}

\usepackage{color}
\usepackage{soul}

\begin{document}

%\preprint{APS/123-QED}

\title{Using numerical methods from nonlocal optics to simulate the dynamics of N-body systems in alternative theories of gravity}

\author{Tiago D. Ferreira}
\email{Corresponding author: tiagodsferreira@hotmail.com}
\affiliation{Departamento de Física e Astronomia da Faculdade de Ciências da Universidade do Porto, Rua do Campo Alegre 687, 4169-007 Porto, Portugal}
\affiliation{INESC TEC, Centre of Applied Photonics, Rua do Campo Alegre 687, 4169-007 Porto, Portugal}

\author{Nuno A. Silva}
\affiliation{Departamento de Física e Astronomia da Faculdade de Ciências da Universidade do Porto, Rua do Campo Alegre 687, 4169-007 Porto, Portugal}
\affiliation{INESC TEC, Centre of Applied Photonics, Rua do Campo Alegre 687, 4169-007 Porto, Portugal}
\affiliation{Texas Center for Advanced Computing, University of Texas at Austin, Advanced Computing Building (ACB), J.J. Pickle Research Campus, Building 205 10100 Burnet Road (R8700), Austin, Texas 78758-4497}

\author{O. Bertolami}
\affiliation{Departamento de Física e Astronomia da Faculdade de Ciências da Universidade do Porto, Rua do Campo Alegre 687, 4169-007 Porto, Portugal}
\affiliation{Centro de Física do Porto, R. Campo Alegre, 4169-007, Porto, Portugal}

\author{C. Gomes}
\affiliation{Departamento de Física e Astronomia da Faculdade de Ciências da Universidade do Porto, Rua do Campo Alegre 687, 4169-007 Porto, Portugal}
\affiliation{Centro de Física do Porto, R. Campo Alegre, 4169-007, Porto, Portugal}

\author{A. Guerreiro}
\affiliation{Departamento de Física e Astronomia da Faculdade de Ciências da Universidade do Porto, Rua do Campo Alegre 687, 4169-007 Porto, Portugal}
\affiliation{INESC TEC, Centre of Applied Photonics, Rua do Campo Alegre 687, 4169-007 Porto, Portugal}
\affiliation{Texas Center for Advanced Computing, University of Texas at Austin, Advanced Computing Building (ACB), J.J. Pickle Research Campus, Building 205 10100 Burnet Road (R8700), Austin, Texas 78758-4497}

\date{\today}% It is always \today, today,

\begin{abstract}
The generalized Schrödinger-Newton system of equations with both local and nonlocal nonlinearities is widely used to describe light propagating in nonlinear media under the paraxial approximation. However, its use is not limited to optical systems and can be found to describe a plethora of different physical phenomena, for example, dark matter or alternative theories for gravity. Thus, the numerical solvers developed for studying light propagating under this model can be adapted to address these other phenomena. Indeed, in this work we report the development of a solver for the HiLight simulations platform based on GPGPU supercomputing and the required adaptations for this solver to be used to test the impact of new extensions of the Theory of General Relativity in the dynamics of the systems, in particular those based on theories with non-minimal coupling between curvature and matter. This approach in the study of these new models offers a quick way to validate them since their analytical analysis is difficult. The simulation module, its performance, and some preliminary tests are presented in this paper.

%\begin{description}
%\end{description}

\end{abstract}

%\pacs{42.65.-k, 42.70.Df, 42.65.Sf, 47.37.+q}% PACS, the Physics and Astronomy
\maketitle

%\tableofcontents

\section{\label{sec:level1}Introduction}

The Theory of General Relativity is currently the most accepted model to describe gravity since it can explain most of the current observations. However, there are some connundrums in the theory, namely the dark matter and dark energy problems and the UV completion of the theory \cite{general_relativity_problems}. Hence, alternative theories of gravity have been proposed in the literature. One of such models is the non-minimal matter-curvature coupling, where,  in addition to a replacement of the scalar curvature by a generic function of it, the matter sector is non-minimally coupled to a further generic function of the Ricci scalar \cite{non_minimal_0}. Proposing such modifications requires that these models have to pass through a series of tests to be considered as valid candidates \cite{nonminimal_2,non_minimal_0,nonminimal_3,nonmininal_1,f_r_functions}, see for example Ref. [\citenum{aritigo_testes}] for a review. However, as already happens with General Relativity, the analytic treatment of these new hypotheses is extremely difficult due to the complexity of the models and, even though numerical solvers can be used, they normally require huge amounts of computational resources. Fortunately, there are certain situations where some approximations reduce the original mathematical model to a general Schrödinger-Newton system where, although the analytical treatment remains difficult, one can use the vast numerical models developed over the years to study and simulate this system. In particular, those used in the context of optics where this model is widely used to describe light propagating in nonlocal optical systems \cite{nematicons,meu_artigo, boson_stars,dark_matter}.

Light propagating in nonlinear and nonlocal systems, under the paraxial approximation, is described by a Schrödinger-Newton system, which consists in a Schrödinger equation coupled with a Newtonian-like potential (or, more generally, a Poisson potential), where the former is responsible for describing the evolution of the envelope of the beam through the medium, and the last one for describing the distribution of the refractive index, and examples of materials where this model is used are nematic liquid crystals \cite{meu_artigo,nematicons,liquid_crystals_book}, thermo-optical materials \cite{boson_stars,dark_matter} and quantum gases \cite{rydberg_gases,quantum_gases_1}. Indeed, the Schrödinger-Newton model is capable of describing a much wider set of systems, ranging from boson stars \cite{boson_stars} to dark matter \cite{dark_matter} and superfluidity \cite{meu_artigo,nonlocal_photon_fluids}, to name a few, and due to the similarities between these mathematical descriptions, this system has been proposed and used for implementing optical analogues \cite{meu_artigo,boson_stars,dark_matter,nonlocal_photon_fluids}, where systems that are hard or even impossible to study are emulated in the laboratory under controlled conditions. Over the years, many numerical models have been developed and improved to simulate this class of systems and in the last years, at our research group, we have developed a set of high-performance solvers based on GPGPU supercomputing to numerically study this class of systems, and successfully applied them in the study of superfluidity in nematic liquid crystals \cite{meu_artigo} and persistent currents in atomic gases \cite{artigo_nuno}. Due to generality of our implementation, these solvers can be applied to other systems and in this work we explore how they can be used to study these alternatives to gravity. In particular, we began by showing how the Schrödinger-Newton system can be used to describe light propagating in different nonlocal systems, and then the approximations required to reduce the original non-minimal coupling models to this mathematical description. The implementation, tests and performance are also discussed.

\section{Physical Model}
The general form of the Schrödinger-Newton system is given by the following set of equations
\begin{equation}
i\frac{\partial\psi}{\partial \tau} = A\nabla^2\psi+B\phi\psi+G\left(|\psi|^2\right)\psi ,
\label{eq:GSNS_1}
\end{equation}
\begin{equation}
\nabla^2\phi+C\phi=D|\psi|^2,
\label{eq:GSNS_2}
\end{equation} 
where the field $\psi=\psi\left(\vec{x},\tau\right)$ represents the wave function of the system under study, such as the optical or mass density for optical or gravitational systems, respectively. The function $G\left(|\psi|^2\right)$ accounts for the possibility of the system having different local nonlinearities, the $\phi=\phi\left(\vec{x},\tau\right)$ accounts for the nonlocal character of the system and is governed by a general Newtonian-like potential given by equation \eqref{eq:GSNS_2}, and $A$, $B$, $C$ and $D$ are constants. The parameter $\tau$ has different interpretations according to the system: for example, in optical systems $\tau$ has dimensions of distance, in gravitational systems, Bose-Einstein condensates or in polariton fluids it has dimensions of time. Equations \eqref{eq:GSNS_1} and \eqref{eq:GSNS_2} can be further generalized if one considers multiple wave functions coupled through the different nonlinearities of the system, however for the present work this generalization is unnecessary.

\subsection{The Schrödinger-Newton equations in optical systems}
As mentioned before, the Schrödinger-Newton system can be used to describe the propagation of an optical beam through a nonlinear medium. If we consider equation \eqref{eq:GSNS_1} alone, we have a mathematical description of the propagation of an envelope $E$ of a light beam, under the paraxial approximation, through an optical medium
\begin{equation}
ik\frac{\partial E}{\partial z}+\frac{1}{2}\nabla^2_{\bot}E+k_0k\Delta nE=0,
\label{eq:equation_to}
\end{equation}  
where $k_0$ and $k=n_0k_0$ are the wavenumbers in the vacuum and inside the medium, respectively. The operator $\nabla^2_{\bot}$ is the two-dimensional Laplacian applied in the two dimensions transverse to the direction of propagation. The refractive index distribution in the transverse plane is given by $\Delta n$. In the case of a medium with local $\left( G\left(|\psi|^2\right)\right) $, and nonlocal $\left(\phi \right) $ nonlinearities, we have that $\Delta n\propto\phi+G\left(|\psi|^2\right)$, where the nonlocal field is governed by equation \eqref{eq:GSNS_2}. For example, in Kerr materials, the local nonlinearity is $G\left(|\psi|^2\right)\propto|\psi|^2$, but higher orders can be considered \cite{cubic_quintic_material}, as well as other types, see Ref. [\citenum{tese_nuno}]. In thermo-optical materials \cite{boson_stars,dark_matter,nonlocal_photon_fluids} the nonlocal field is governed by
\begin{equation}
\nabla_{\bot}^2\Delta n -\frac{1}{\sigma^2}\Delta n= -\frac{\zeta\mu}{\kappa}|E|^2,
\label{eq:to_nonlocal_index_change_corrected}
\end{equation}
\newcommand{\soom}{\sim}where $\sigma$ is the nonlocal interaction length scale ($\sigma\soom d/2$, with $d$ corresponding to the diameter of the medium \cite{boson_stars}), $\mu$ and $\zeta$ are the thermo-optical and linear absorption coefficients, respectively, and $\kappa$ is the thermal conductivity. For nematic liquid crystals \cite{meu_artigo,nematicons,NLC_playground} we have that
\begin{equation}
\nu\nabla^2\theta-2q\theta=-2|E|^2,
\label{eq:full_NLC_normalized_theta}
\end{equation}
where $\theta$ represents the perturbation induced in the molecular director due to the electric field and $\theta\propto\phi$, $q$ is related to the pre-tilt of the molecular director and $\nu$ is the normalized elastic coefficient.

\subsection{The Schrödinger-Newton for Non-Minimal coupling models}

The idea behind alternative models of gravity is to modify the Einstein-Hilbert action to account for several phenomena that are not explained by General Relativity. In particular, the non-minimal matter-curvature coupling model consists in replacing the Ricci scalar or curvature scalar, $R$, with a general function of it $f_1(R)$, and coupling to the matter Lagrangian density with the curvature through another general function, $f_2(R)$ \cite{non_minimal_0}. These modifications lead to 
\begin{equation}
S=\int d^4x\sqrt{-g}\left[ \frac{1}{2}f_1\left(R\right)+f_2\left(R\right)\mathcal{L}_m\right],
\label{eq:non_minimal_action} 
\end{equation}
where $g$ is the determinant of the metric, $g_{\mu\nu}$. In the limits $f_1(R)\rightarrow R$ and $f_2(R)\rightarrow 1$, in units such that $M_p^2\equiv(8\pi G)^{-1}=1$, General Relativity is recovered. In the Newtonian limit, this model predicts that the usual gravitational potential has to be corrected to  
\begin{equation}
\phi = \phi_N + \phi_C,
\end{equation}
where $\phi_N$ is the Newtonian potential, which obeys a modified Poisson equation
\begin{equation}
\left(3\alpha\nabla^4-\nabla^2\right)\phi_N=\left(4\alpha-\beta\right)\nabla^2\rho-\gamma\rho,
\label{eq:full_phi_potential}  
\end{equation}
where $\alpha=f_1''(0)/f_1'(0)$ is a constant that measures deviations from General Relativity in the pure gravity sector and the primer denotes derivations with respect to $R$, and $\gamma=f_2(0)/f_1'(0)$ and $\beta=f_2'(0)/f_1'(0)$ can be parameterized in the following form
\begin{equation}
\gamma\equiv \begin{cases}
\frac{1}{2}, & \text{General Relativity}\\
\frac{1}{2}\left[1-\left(\frac{r_0}{r+r_0}\right)^{\lambda}\right], & \parbox{3cm}{\centering Non-Minimal Coupling Models}
\end{cases},
\end{equation}
\begin{equation}
\beta\equiv\lambda\left(\frac{r_0}{r+r_0}\right)^{\lambda}\frac{1}{\left(r+r_0\right)},
\end{equation}
where $\lambda$ measures the strength of the Non-Minimal coupling and $r_0=M/4\pi$ corresponds to the Schwarzschild radius. The potential $\phi_C$ is due to the non-minimal coupling and reads \cite{orfeu_pure_coupling} 
\begin{equation}
\phi_C=\ln\left[1-\left(\frac{r_0}{r+r_0}\right)^{\lambda}\right].
\label{eq:phi_c}
\end{equation}

Let us consider a pure Non-Minimal coupling ($\alpha=0$), which is translated into
\begin{equation}
\nabla^2\phi_N=\beta\nabla^2\rho+\gamma\rho.
\label{eq:full_phi_potential_alpha_zero}  
\end{equation}

At large scales, matter can be modeled as a perfect fluid \cite{gravity_hidro}, hence, from the full Vlasov-Poisson system, it is possible to obtain a set of hydrodynamic equations capable of describing the dynamics of the density $\rho$ and velocity $\vec{v}$ of the fluid
\begin{equation}
\frac{\partial \rho}{\partial t} + \nabla\cdot(\rho\vec{v})=0
\label{eq:cosmology_hydro_1}
\end{equation}
\begin{equation}
\frac{\partial\vec{v}}{\partial t} + (\vec{v}\cdot\nabla)\vec{v}+\frac{\nabla P}{\rho} + \nabla\phi=0,
\label{eq:cosmology_hydro_2}
\end{equation}
where we have assumed a polytropic relation of the form $P=w\rho^n$, with $w$ and $n$ being coefficients to be chosen for each physical situation. Equations \eqref{eq:cosmology_hydro_1} and \eqref{eq:cosmology_hydro_2} correspond to the continuity and Euler equations, respectively. The benefit in beginning with this hydrodynamic approach is that, by applying the Madelung transformation \cite{madelung}, this set of equations is transformed into a Schrödinger-Newton system. This correspondence is normally used to establish a hydrodynamic analogy for quantum mechanics \cite{gravity_hidro}, however, in this work we use this approach in the opposite direction. This transformation is achieved by writing: $\psi=\sqrt{\rho}e^{i\Phi/\nu}$, where $\nu$ work as an analogous of the reduced Planck constant, $\hbar$, $\rho=|\psi|^2$ and $\vec{v}=\nabla\Phi$. In addition, plugging these redefinitions into in equations \eqref{eq:cosmology_hydro_1} and \eqref{eq:cosmology_hydro_2}, it follows
\begin{equation}
i\nu\frac{\partial\psi}{\partial t}=-\frac{\nu^2}{2}\nabla^2\psi+\left[\nu^2V_B +\phi_N +\phi_C +V_{p}\right]\psi
\label{eq:SN_cosmology}
\end{equation}
where $V_B=\nabla^2\sqrt{\rho}/(2\sqrt{\rho})$ is the Bohm quantum potential and $V_{pressure}$ is given by
\begin{equation}
V_{pressure}=\begin{cases}
w\ln(\rho) & n=1\\
\frac{nw}{n-1}\rho^{n-1} & n>1 \lor n<0
\end{cases}.
\end{equation}

Through this set of transformations and approximations, we obtain a set of equations that corresponds to a general form of the Schrödinger-Newton system. Hence, it can be solved by adapting the solvers initially developed for equations \eqref{eq:GSNS_1} and \eqref{eq:GSNS_2}.

\section{Implementation and Performance analysis}

The original solvers developed for simulating the optical systems were implemented with a modular structure, with an hardware-neutral paradigm, and were based on GPGPU supercomputing. These solvers are integrated into a more general one, HighLight, which offers the possibility of simulating different light-matter systems with different interactions and levels of approximation. Analyzing, in particular, the Schrödinger-Newton solver, this was implemented using standard beam propagation methods, where the Schrödinger equation \eqref{eq:GSNS_1} is solved with the Symmetric Split-Step Fourier Method since it is the most efficient one \cite{SSFM_NSE} and takes advantage of the well-optimized Fourier methods implemented for GPGPU. To maintain consistency and efficiency, equation \eqref{eq:GSNS_2} is also solved with spectral methods. All the solvers were written in C++ and the GPGPU implementation with an hardware-neutral approach was done through the ArrayFire library \cite{arrayfire}. The implementation was initially tested, and the works \cite{meu_artigo,artigo_nuno} already published based on these solvers assures its physical correctness. After performing the required adaptations to solve equation \eqref{eq:SN_cosmology}, we tested the solver performance when running with different hardware platforms, GPUs and CPUs, as well as with different Application Programming Interfaces (APIs) (integrated with the Arrayfire), namely CUDA and OpenCL.

\begin{figure}[!ht]
	\centering
	\includegraphics[width=0.45\textwidth]{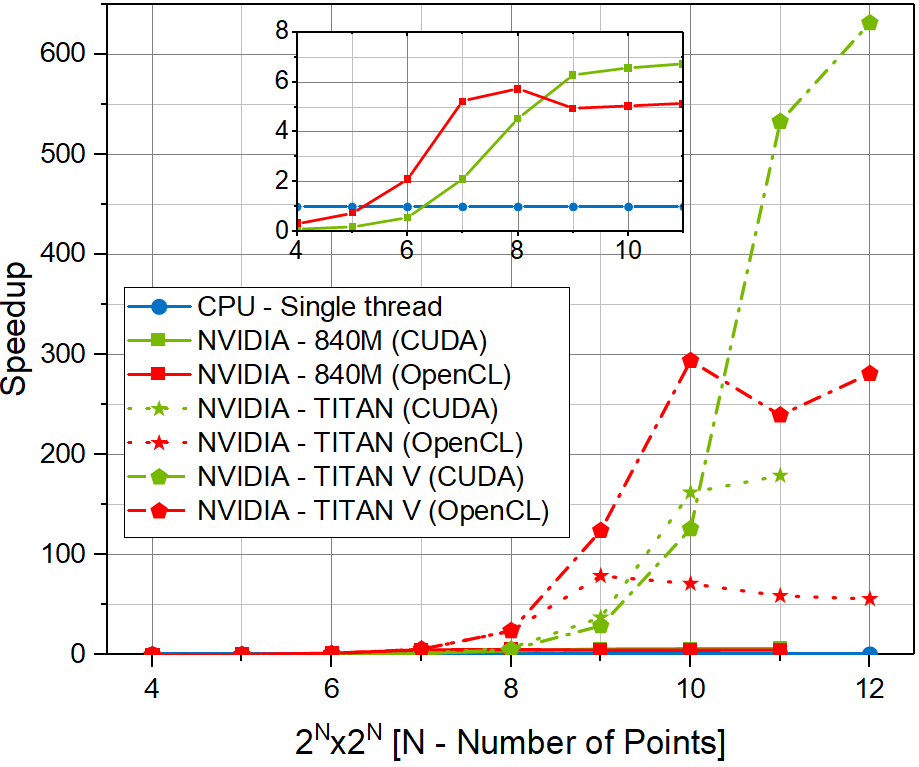}
	\caption{Speedup comparison for different hardware and software configurations with an high-end CPU in single-thread. The insets are a zoom of the NVIDIA 840M (laptop GPGPU) results.}
	\label{fig:bench_results_single}
\end{figure}
\begin{figure}[!ht]
	\centering
	\includegraphics[width=0.45\textwidth]{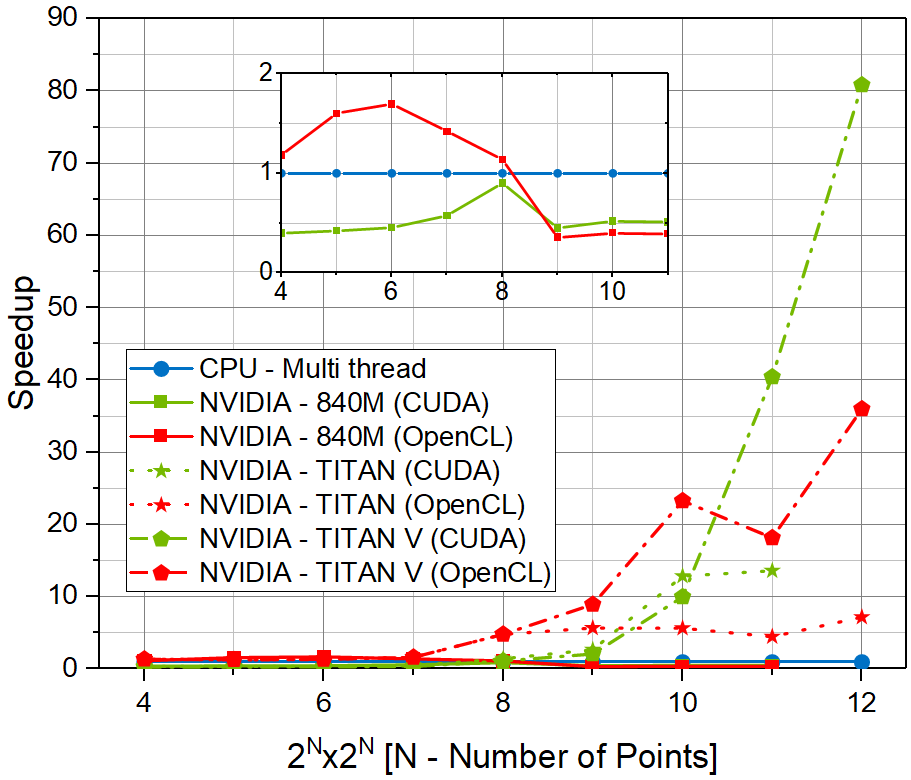}
	\caption{Speedup comparison for different hardware and software configurations with an high-end CPU in Multi-thread. The insets are a zoom of the NVIDIA 840M (laptop GPGPU) results.}
	\label{fig:bench_results_multi}
\end{figure}

Figures \ref{fig:bench_results_single} and \ref{fig:bench_results_multi} present the speedup results against a high-end CPU running in single and multi thread mode, respectively, in double precision (corresponding to a number representation of 64 bits), and for large mesh points the GPGPUs outperform CPUs. In particular, with our most powerful GPGPU (NVIDIA - TITAN V), the solver can be more than $600$ or $80$ times faster when compared with the CPU in single or multi thread mode, respectively. Even with a GPGPU from a personal laptop (NVIDIA 840M), the speedup results can be greater than $6.5$. In this configuration, and when we compare with the CPU in multi thread mode, the laptop is slower, however, we have to remember that we are comparing with a powerful CPU, and against the laptop CPU, the GPGPU is much faster. Furthermore, the speedup results also show that the optimal hardware-software configuration depends on the mesh size of the problem.

Lastly, the results demonstrate the advantage of using a GPGPU based implementation. With this adapted solver we are now able to perform numerical simulations of cosmological systems under different models of gravity with reasonable simulation times.

\section{Results}

\begin{figure}[!ht]
	\centering
	\includegraphics[width=0.48\textwidth]{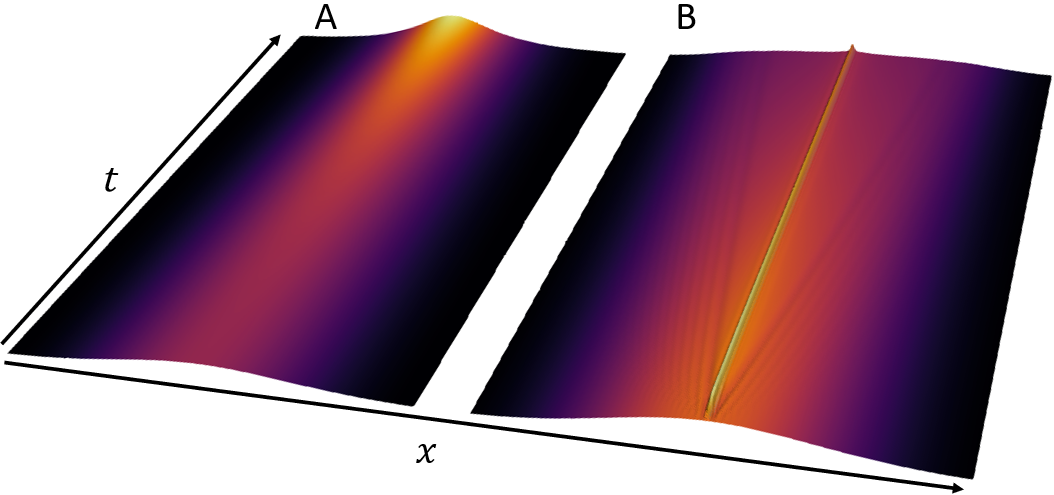}
	\caption{Slices of a two dimensional simulation over the time evolution for an initial Gaussian mass distribution in the original (\textbf{A}) and modified (\textbf{B}) version of the Theory of General Relativity. In these simulations the parameters $M_0=25$, $\nu=1$, $w=1.02$, $n=1$ and $\lambda=1$ were used. The variables $x$ and $t$ are in arbitrary units.}
	\label{fig:results}
\end{figure}

\begin{figure}[!ht]
	\centering
	\includegraphics[width=0.45\textwidth]{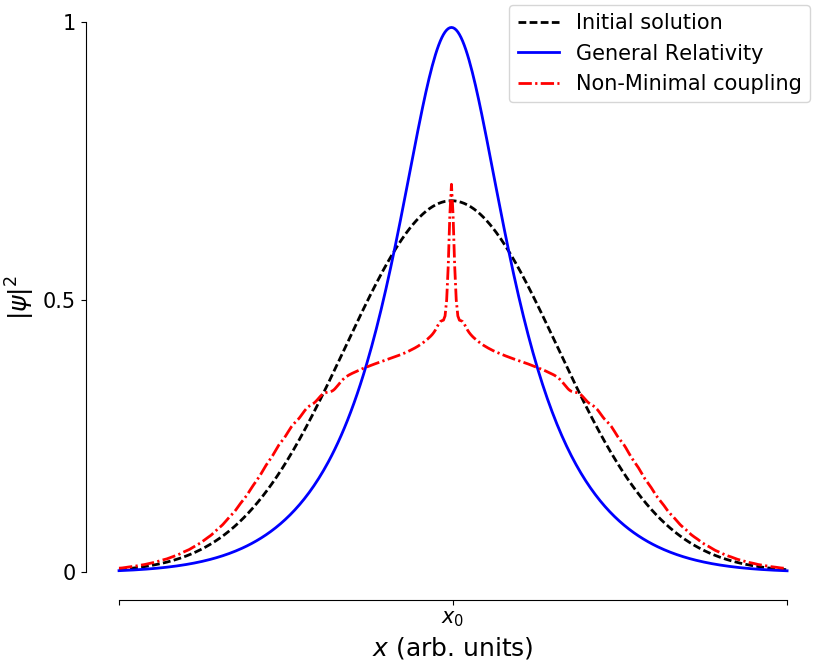}
	\caption{Normalized initial and final profiles of Figure \ref{fig:results}.}
	\label{fig:results_profile}
\end{figure}

The modified solver was then used to evaluate the differences in the evolution of a distribution of matter under different gravitational models. In every case, the simulation was initiated by considering a Gaussian distribution of the number density of massive particles, sufficiently dense to constitute a fluid. Using the optical analogue model generated by the standard General Relativity and a Non-Minimal coupling model, as described by equation \eqref{eq:SN_cosmology}, we allowed the number density distribution to evolve. The results are shown in Figures \ref{fig:results} and \ref{fig:results_profile}. Even though the two models start from the same initial distribution, their distinct gravitational character results in completely different processes of self-organization. Accordingly to the results predicted by General Relativity, the initial distribution as a whole slowly collapses at the center of the Gaussian. On the other hand, the results predicted for the Non-minimal coupling reveal two distinct dynamical processes: part of the distribution collapses at the center of the initial distribution, after which appears to stabilize. As a result of this rapid accumulation of mass, are formed shock waves that can be seen in Figure \ref{fig:results}. Meanwhile, another component of the initial distribution appears to escape the gravitational well at the center, as it overcomes the escape velocity.  

These results seem to complement the analytic analysis of Ref.  [\citenum{orfeu_pure_coupling}], where it was shown that the choice/parameterization, equation \ref{eq:phi_c}, leads to a stronger gravitational pull due to the fact that the non-minimal coupling is acting as an effective pressure.

Figure \ref{fig:results_profile} shows the initial and final profiles of number density distribution. Clearly, the two models of gravity produce distinct outcomes thus showing the usefulness of this method as test-bed to investigate alternative models of gravity.

\section{Conclusions}

This article describes how solvers initially developed to simulate the propagation of light in nonlinear and nonlocal optical media under the paraxial approximation can be adapted to investigate the dynamics of matter generated by different models of gravitational interaction. 

In detail, we have demonstrated that for non-minimal coupling theories of gravity under certain approximations the evolution equations of matter coincide with those used to model light propagation. As a result, the study of extensions of the Theory of General Relativity and their implications in processes of self-organization of matter can benefit from the development of high performance tools in optics based on GPGPU technologies. These technologies allow for acceleration factors up to $600$ in calculation times when compared with the conventional approach based on CPUs, while permitting to simulate problems with high spatial and temporal resolution. Thus, these solvers provide an efficient test-bed to evaluate non-minimal coupling models against observations of the dynamics of astronomical objects. At the same time, these solvers bridge nonlinear optics and gravity, suggesting the possibility to emulate gravitational dynamics produced by different models in optical experiments.

\section*{Acknowledgments}
This work is supported by the ERDF – European Regional Development Fund through the Operational Programme for Competitiveness and Internationalisation - COMPETE 2020 Programme and by National Funds through the Portuguese funding agency, FCT - Fundação para a Ciência e a Tecnologia within project $\ll$POCI-01-0145-FEDER-032257$\gg$

\bibliographystyle{apsrev4-1}
%\bibliographystyle{abbrv}
%\bibliography{referencesCosmo}
%\bibliographystyle{aipauth4-1}



\end{document}